# Unveiling neuronal microstructure in the human brain *in vivo* with time-dependent radial diffusivity in MRI

Ante Zhu[1], Jerome J. Maller[2], H. Douglas Morris[3,4], Jeffrey P. Guenette[5], Flavio Dell' Acqua[6], Chu-Yu Lee[2], Jeffrey McGovern[7], Stephan E. Maier[5], Raymond Y. Huang[5], Andrew L. Alexander[8], Maureen N. Hood[3,4], Desmond Teck Beng Yeo[1], Steven CR Williams[9], Carl-Fredrik Westin[5], Vincent B. Ho[3,4], Vincent A. Magnotta[2], Hua Li[7]

[1] Technology & Innovation Center, GE HealthCare, Niskayuna, NY, USA;

[2] Department of Radiology, The University of Iowa, Iowa City, IO, USA;

[3] Uniformed Services University, Bethesda, MD, USA;

[4] Walter Reed National Military Medical Center, Bethesda, MD, USA;

[5] Department of Radiology, Mass General Brigham, Harvard Medical School, Boston, MA, USA;

[6] Neuroimaging and Biophysics Lab, Department of Neuroimaging, Institute of Psychiatry, Psychology and Neuroscience, King's College London, London, UK

[7] MRI Engineering, GE HealthCare, Waukesha, WI, USA;

[8] Waisman Center, University of Wisconsin-Madison, Madison, WI, USA;

[9] Department of Neuroimaging, Institute of Psychiatry, Psychology and Neuroscience, King's College London, London, UK

**Correspondence**:

Ante Zhu, Address: One Research Circle, MR 135, Niskayuna, NY, USA; Email: ante.zhu@gehealthcare.com; Tel: +15186884603.

**Abstract**

Diffusion time-dependence, defined as variations in diffusivity and/or diffusional kurtosis with diffusion time, has emerged as a valuable non-invasive imaging marker for characterizing tissue microstructural features, such as cell size, density, packing disorder, and membrane permeability. In white matter, diffusion time-dependent changes between the short diffusion time and long diffusion time in radial diffusivity (RD), defined as the diffusivity perpendicular to fiber tracts, were demonstrated to correlate strongly with mean axon diameter in *ex vivo* spinal cord tissues, and to reveal demyelination in mouse corpus callosum. Despite their potential to non-invasively unveil neuronal microstructures to improve the assessment and targeted therapy of neurological diseases, these novel image contrasts obtained at short diffusion times using oscillating gradient spin echo (OGSE) have only recently become feasible for human *in vivo* studies with high-performance gradient MRI systems. In this preliminary study, we characterized time-dependent RD with OGSE encoding in the human brain *in vivo*. The change in radial diffusivity between short diffusion time and long diffusion time (∆RD) consistently exhibited high values in the corticospinal tract, indicating high sensitivity of ∆RD to large axon diameter in human brains. Imaging at a high OGSE frequency of 100 Hz and a moderate b-value of 800 s/mm$^2$ produced the highest ∆RD in the corticospinal tract. This study established a baseline for future investigations of neuronal microstructural alterations in neurological disorders and diseases.

**Keywords**: neuronal microstructure; time-dependent diffusion; oscillating gradient spin echo; radial diffusivity; axonal diameter; human high-performance gradient.

## 1. Introduction

Neuronal microstructural features, including axon diameter, density, orientation, myelination and extracellular composition, form the structural basis of brain connectivity and function. Accurately characterizing these features can have significant clinical impact in patients with neurologic impairment. For example, axon diameter influences nerve conduction velocity, with larger axons enabling faster signal transmission and more efficient long-range connectivity(Bin et al., 2024; Seidl, 2014), as exemplified by the corticospinal tract. Monitoring changes in axon diameter may provide insight into axonal injury and/or degeneration. The ability to identify early diffuse axonal injury, which is typically associated with traumatic brain injury, can provide invaluable information for treatment planning and surveillance of patient response to therapy. Additionally, demyelination and remyelination are hallmark features of numerous neurodegenerative and demyelinating conditions, such as multiple sclerosis, Alzheimer's disease, vascular cognitive impairment, and long-COVID-related or post-infectious cognitive dysfunction. These conditions would similarly benefit from the ability to detect early microstructural alterations.

Time-dependent diffusion MRI has emerged as a powerful noninvasive technique for characterizing tissue microstructure and detecting alterations that may be invisible or nonspecific with conventional anatomical imaging techniques(Aggarwal et al., 2012; Aggarwal et al., 2020; Wu et al., 2014). Changes in diffusivity and diffusional kurtosis across diffusion times reflect the interaction between water molecular motion and cellular structures. Diffusion time-dependence has been shown to correlate with cell size(Xu et al., 2021), density(Wu et al., 2021), and membrane permeability(Li et al., 2023; Olesen et al., 2022)—key pathological biomarkers that currently require biopsy and histopathology analysis. Oscillating gradient spin echo (OGSE) diffusion encoding(Zhu et al., 2025) enables probing of short diffusion time ≤10 ms, providing sensitivity to microstructural features over the short length scale ≤10 μm, which are particularly relevant to neuronal architecture in the human brain. In comparison, conventional pulsed gradient spin echo (PGSE) diffusion encoding achieves relatively long diffusion times >10 ms.

In simulations and preclinical studies, diffusivity perpendicular to fiber tracts (i.e., radial diffusivity, RD) has been shown to exhibit diffusion time-dependence from long diffusion times to short diffusion times(Alderson et al., 2025; Harkins et al., 2021; Xu et al., 2023). Simulation studies by Harkins, et al. (Harkins et al., 2021) and Tetreault, et al.(Tetreault et al., 2020) showed that ΔRD—the estimated change in RD between short and long diffusion times—increased monotonically within a range of effective mean axon diameter ≥2 μm. In comparison, RD at either short diffusion time or long diffusion time were dominated by extra-axonal water(Tetreault et al., 2020). The same study has demonstrated a strong linear correlation between ΔRD and area-weighted mean axon diameter at ≥2 μm derived from scanning electron microscopy of *ex vivo* ferret spinal cord tissues. Collectively, these findings support the potential of a simple estimate of ΔRD as a surrogate marker of axon diameter.

Despite its promise for noninvasive tissue microstructure characterization, time-dependent diffusion MRI with OGSE encoding has only recently been translated to human *in vivo* applications, largely enabled by the advent of human high-performance gradient MRI systems. Zhu, et al.(Zhu et al., 2025) articulated that achieving OGSE diffusion encoding at short diffusion times (corresponding to high OGSE frequencies), high b-values, and short echo times requires simultaneously high gradient amplitude, high slew rate, and an elevated peripheral nerve stimulation (PNS) threshold in humans. These requirements are satisfied by head-only high-performance gradient MRI systems(Feinberg et al., 2023; Foo et al., 2020; Ramos-Llordén et al., 2025; Rosler et al., 2020) that operate at maximum gradient amplitude of 200-500 mT/m, slew rate of 750-900 T/m/s, and increased PNS limits. Using these systems, high-quality OGSE diffusion MRI has been successfully demonstrated in the human brain *in vivo*(Dai et al., 2023; Lan et al., 2026; Michael et al., 2022; Zhu et al., 2023). Furthermore, in brain tumor imaging, Zhu et al.(Zhu et al., 2023) leveraged the ratio of diffusivity measured at short and long diffusion times to probe tumor cellular density, indicating the potential of time-dependent diffusion for more accurate evaluation of tumor aggressiveness, infiltration, and recurrence. The success of these preliminary studies substantially strengthens confidence in the feasibility of *in vivo* tissue microstructure characterization on high-performance human gradient MRI systems.

This study aimed to characterize the time-dependent radial diffusivity at short and long diffusion times in human white matter and to investigate its potential in characterizing axon diameter. We further assessed the sensitivity of different imaging protocols (e.g., OGSE frequencies, b-values, and TE’s) to neuronal microstructures in human brain on a head-only gradient MAGNUS 3.0T MRI system (Foo et al., 2020) at maximum performance of 300 mT/m and 750 T/m/s.

## 2. Methods

### 2.1. Study Population

This prospective study was approved by the local Institutional Review Board. Healthy adults were recruited. All participants provided written informed consent prior to MRI scans.

### 2.2. MRI acquisitions

MRI scanning was performed in an investigational head-only MAGNUS 3.0T MRI system (GE HealthCare, Waukesha, WI, USA) operating at maximum performance of 300 mT/m and 750 T/m/s at 2 MVA input power per gradient axis. A 32-channel radiofrequency receive coil array (Nova Medical, Wilmington, MA, USA) was used for signal reception. $T_1$-weighted MPRAGE anatomical images at 1 mm isotropic resolution were acquired. Diffusion MRI with PGSE and OGSE encoding were applied to obtain time-dependent diffusion tensor imaging (DTI). Cosine-modulated trapezoid OGSE waveforms(Van et al., 2014) within PNS limits were used. 2D single-shot echo-planar imaging (EPI) was applied for k-space encoding. A $T_2$-weighted EPI with reversed phase-encoding gradient polarity was also acquired for "blip-up/blip-down" distortion correction(Holland et al., 2010).

Five time-dependent DTI acquisitions were obtained with different diffusion times, b-values, and TE's, as summarized in **Table 1**. The rationale of the protocol design is as follows:

(1) "High-frequency Medium-b Long-TE" at OGSE frequency of 100 Hz and b-value of 800 s/mm$^2$. This is a combination of the highest OGSE frequency with an adequate b-value of 800 s/mm$^2$ achievable at TE $\leq$125 ms.

(2) "Medium-frequency High-b Long-TE" at OGSE frequency of 85 Hz and b-value of 1000 s/mm$^2$. b-value of 1000 s/mm$^2$ is the common clinical diffusion weighting for diffusivity measurement in the human brain *in vivo*. The OGSE frequency of 85 Hz was determined as the maximum frequency at a minimum b-value of 1000 s/mm$^2$ with TE $\leq$120 ms.

(3) "Medium-frequency Low-b Short-TE" at OGSE frequency of 85 Hz and b-value of 500 s/mm$^2$. Typically, b-values $\geq$1000 s/mm$^2$ in PGSE encoding are considered more sensitive to restricted diffusion in white matter. In OGSE encoding at short diffusion times, the measured diffusivity is larger compared to that in PGSE encoding at long diffusion time. Thus, a lower b-value in OGSE encoding may be sensitive to restricted diffusion, as shown in tumor applications by Zhu et al(Zhu et al., 2023). In addition, lower b-values significantly reduce TE(Zhu et al., 2025) and boost signal-to-noise ratio (SNR). Therefore, we included this acquisition to assess the sensitivity of medium-frequency low-b acquisitions in human white matter.

(4) "Low-frequency High-b Short-TE" at OGSE frequency of 54 Hz and b-value of 1000 s/mm$^2$. The OGSE frequency of 54 Hz was determined as the lowest frequency at a b-value of 1000 s/mm$^2$ at the shortest TE. Lower OGSE frequencies result in a longer TE due to the minimal diffusion encoding time, i.e., one oscillating period $1/f$ ($f$ is the OGSE frequency). Higher OGSE frequencies require more oscillating periods to achieve a high b-value, which also results in long diffusion encoding time and TE(Zhu et al., 2025).

(5) "Low-frequency Low-b Long-TE" at OGSE frequency of 28 Hz and b-value of 500 s/mm$^2$. We further evaluated the time-dependent radial diffusivity measurement achievable at lower gradient performance, such as in clinical whole-body MRI systems operating at maximum performance of 80 mT/m and 200 T/m/s. The limited gradient performance results in low b-value and long TE at relatively low OGSE frequencies.

In each of the five time-dependent DTI acquisitions, a PGSE encoding was also obtained in addition to the OGSE encoding acquisition, with the matched spatial coverage, spatial resolution, b-value, TE, TR, and diffusion encoding directions. The pulse sequence timing diagrams for both OGSE and PGSE are shown in the **Figure 1**. In PGSE encoding, a mixing time in the range of 38-103 ms and a small delta (i.e., the time duration from the beginning of the ramp to the end of the gradient plateau) of 2-8 ms were obtained.

### 2.3. MRI data processing and analysis

Diffusion MRI datasets were reconstructed using a customized deep-learning phase correction technique (Chien et al., 2025; Cochran et al., 2026) to reduce the noise floor. A commercialized deep-learning model (Lebel, 2020) was further applied for denoising and Gibbs ringing reduction. EPI image distortion was corrected by applying a "blip-up/blip-down" correction(Holland et al., 2010). Residual eddy current-induced image distortion from the diffusion encoding gradient and/or motion were corrected by registering each diffusion-weighted image to $T_2$-weighted image (i.e., b=0 s/mm$^2$) using a customized tool (Bhushan et al., 2015; Bhushan et al., 2012). All DTI images were spatially co-registered to the $T_1$-weighted MPRAGE. White matter parcels were segmented and registered to a custom brain atlas with 50 white matter parcels(Mori et al., 2008).

The diffusion metrics, mean/axial/radial diffusivities (MD/AD/RD) and fractional anisotropy (FA) maps, at each diffusion time were calculated after correcting for gradient nonlinearity-induced spatially varying b-value and diffusion tensor on diffusion encoding following an established method(Tan et al., 2013). The difference between RD at two diffusion times was calculated as ΔRD=RD(OGSE $f$, b-value)-RD(PGSE $t_d$, b-value), where $f$ is the OGSE encoding frequency and $t_d$ is the mixing time. Therefore, it resulted in five ΔRD, i.e., ΔRD(100 Hz, b800)=RD(100 Hz, b800)-RD(103 ms, b800), ΔRD(85Hz, b1000)=RD(85 Hz, b1000)-RD(96 ms, b1000), ΔRD(85 Hz, b500)=RD(85 Hz, b500)-RD(49 ms, b500), ΔRD(54Hz, b1000)=RD(54 Hz, b1000)-RD(38 ms, b1000), and ΔRD(28 Hz, b500)=RD(28 Hz, b500)-RD(68 ms, b500).

In each of the 50 white matter parcels, the mean and standard deviation of RD(PGSE), RD(OGSE), and ΔRD were calculated.

## 3. Results

A total of four volunteers (3 female, 1 male; age 55–70 years) were recruited. Representative MD/AD/RD maps measured from OGSE 100 Hz and PGSE, as well as ΔMD/ΔAD/ΔRD maps, from the "High-frequency Medium-b Long-TE" protocol in a single subject are shown in **Figure 2**. MD, AD, and RD measured from OGSE encoding all showed higher values in white matter, compared to those measured from PGSE encoding, as indicated by the positive ΔMD/ΔAD/ΔRD values. Both ΔMD and ΔRD maps revealed strikingly elevated values concentrated in the corticospinal tract, while most other white matter regions showed near-zero or modest ΔMD and ΔRD.

Representative ΔRD maps from all five time-dependent DTI acquisitions are shown in **Figure 3**. The corticospinal tract, especially in the posterior limbs of the internal capsules, showed hyper-intensities (i.e., higher ΔRD values compared to other brain tissues) in the ΔRD(100 Hz, b800), ΔRD(85 Hz, b1000), ΔRD(85 Hz, b500) and ΔRD(54 Hz, b1000) maps. In comparison, the corticospinal tract did not show hyper-intensities in the ΔRD(28 Hz, b500). Among all five acquisitions, ΔRD(100 Hz, b800) produced the highest ΔRD compared to the other four acquisitions with lower OGSE frequencies (OGSE 28–85 Hz).

Multiple slices of ΔRD(85 Hz, b1000) maps in two female volunteers at similar ages (55 years old and 57 years old), as well as $T_1$-weighted MPRAGE images, are shown in **Figure 4**. In Subject #3 (**Figures 4A and 4C**), ΔRD of the white matter fibers of the major motor pathways, including corona radiata (as indicated by white arrows in **Figure 4A**) and internal capsule (as indicated by blue arrows in **Figure 4A**), exhibited hyper-intensities. This spatial pattern was consistently observed across Subject #1 (**Figure 2**) and Subject #2 (**Figure 3**) as well. In Subject #4 (**Figure 4B and 4D**), hypo-intensities in $T_1$-weighted MPRAGE were present in the left hemisphere, as indicated by red arrows. The subject had a prior middle cerebral artery stroke. Compared to the other three healthy subjects (e.g., Subject #3 in **Figure 4A**), the corona radiata on both hemispheres of Subject #4 did not exhibit distinctly high ΔRD (as indicated by the white arrows in **Figure 4B**). The internal capsule on both hemispheres of Subject #4 exhibited smaller regions with hyperintense ΔRD (indicated by the blue arrows in **Figure 4B**).

Multiple slices of ΔRD(54 Hz, b1000) maps in the two female volunteers shown in **Figure 4** (i.e., Subjects #3 and #4) are presented in **Figure 5**. In Subject #3, ΔRD(54 Hz, b1000) of the corona radiata (as indicated by the white arrows in **Figure 5A**) and internal capsule (as indicated by the blue arrows in **Figure 5A**) exhibited subtle hyper-intensities, compared to ΔRD(85 Hz, b1000) shown in **Figure 4A**. In Subject #4, ΔRD(54 Hz, b1000) maps also exhibited reduced regions with hyperintense ΔRD in the internal capsule of both hemispheres (as indicated by the blue arrows in **Figure 5B**), similar to ΔRD(85 Hz, b1000) shown in **Figure 4B**.

Mean ΔRD values across all white matter parcels in Subjects #1, #2, and #3 are summarized in **Figure 6**, including ΔRD measured from all five time-dependent DTI protocols. In all three subjects, most white matter parcels exhibited values that decreased progressively from ΔRD(100 Hz, b800), ΔRD(85Hz, b1000), ΔRD(85 Hz, b500), ΔRD(54Hz, b1000), to ΔRD(28 Hz, b500). Among all the white matter parcels, corticospinal tract (atlas-defined CST ROI), inferior cerebellar peduncle (ICP), posterior limb of internal capsule (PLIC), superior corona radiata (SCR), and superior longitudinal fasciculus (SLF) consistently showed high ΔRD in all three subjects. Furthermore, the genu of corpus callosum (gCC) showed lower ΔRD compared to the body (bCC) and the splenium (sCC) in all three subjects.

Quantitatively, this frequency-dependent contrast is substantial (**Table 2**): the all-subject mean ΔRD(100 Hz, b800) in the corticospinal tract (~0.21 μm²/ms, averaged across Subjects #1, #2, and #3) was approximately 2.5-fold higher than in the genu of corpus callosum (0.08 μm²/ms) and roughly 1.7-fold higher than in the anterior corona radiata and splenium (0.12 μm²/ms each) — regions dominated by small-diameter callosal axons (**Table 2**). This contrast narrowed markedly as OGSE frequency decreased: at ΔRD(28 Hz, b500), the corticospinal tract (~0.05 μm²/ms) was only 1.2- to 1.6-fold higher than these same forceps minor and forceps major regions (0.03–0.04 μm²/ms), mirroring the loss of visible hyperintensity in the ΔRD(28 Hz, b500) maps (**Figure 3**) and the convergence of parcel-wise ΔRD values at low OGSE frequency in **Figure 6**. Because ΔRD(100 Hz, b800) requires oscillating gradients delivered at high amplitude and slew rate—achievable only on a head-only high-performance gradient system—this pronounced corticospinal tract contrast could not have been captured with conventional whole-body MRI gradients.

**Figure 7** shows the comparison of ΔRD(85 Hz, b1000) between Subjects #1-3 and Subject #4. As already reflected in the image contrast of the ΔRD(85 Hz, b1000) map in **Figure 4**, Subject #4 exhibited a lower average ΔRD(85 Hz, b1000) in the corticospinal tract, compared to the other three subjects.

## 4. Discussion and Conclusions

We performed an initial study of diffusion time-dependent changes in radial diffusivity (ΔRD) between short and long diffusion times in human white matter in four subjects, using OGSE diffusion encoding with a head-only high-performance gradient MRI system. In three healthy subjects, ΔRD of corticospinal tract regions (e.g., corona radiata, internal capsule, cerebral peduncle, pontine) showed higher values compared to other white matter tracts. Among five time-dependent DTI protocols with different OGSE frequencies, b-values, and TE's, the protocol with the highest OGSE frequency of 100 Hz and a moderate b-value of 800 s/mm$^2$ yielded the highest ΔRD values. Furthermore, in a subject who had a prior middle cerebral artery stroke, lower ΔRD was observed in the key white matter tracts, compared to the other three healthy subjects.

The observation of hyperintense ΔRD in the corticospinal tracts of the three healthy subjects suggested that ΔRD measured at short diffusion times (equivalent to high OGSE frequencies) may preferentially highlight white matter tracts known to contain large-diameter axons. Prior quantifications of axonal diameters, both from histopathology analysis(Duval et al., 2018) and from MRI-based measurements(Fan et al., 2020; Huang et al., 2020; Ma et al., 2026), indicated the presence of large axonal diameters in human corticospinal tracts. In addition, simulations(Harkins et al., 2021; Tetreault et al., 2020) and recent preclinical studies(Alderson et al., 2025) demonstrated a strong linear correlation between ΔRD and effective mean axon diameters when no smaller than 2 $\mu$m. The spatial pattern showing high ΔRD of the corticospinal tract from this study in the human brain *in vivo*, in comparison to that of forceps minor and forceps majors, is consistent with the spatial pattern showing high axon diameter index of the corticospinal tract from PGSE-based axon diameter mapping techniques (e.g., Figure 5 in (Fan et al., 2020) and Figure 2 in (Huang et al., 2020)). Therefore, our study has shown the great potential of using ΔRD at high OGSE frequencies to characterize axonal diameters of the human brain *in vivo*.

Time-dependent DTI acquisitions at high OGSE frequencies with adequate b-values, such as ΔRD(100 Hz, b800) and ΔRD(85 Hz, b1000), showed the highest ΔRD contrast between the corticospinal tract and other white matter regions. This observation is consistent with the expected high sensitivity of shorter diffusion times to restricted diffusion within larger-diameter axons. At a fixed OGSE frequency (e.g., 85 Hz in this study), a higher b-value of 1000 s/mm$^2$ generated greater ΔRD contrast compared to a lower b-value of 500 s/mm$^2$, although high b-value results in substantially longer TE (e.g., 115 ms vs. 68 ms) which leads to a lower SNR and different $T_2$ relaxation effects of intra- and extra- axonal water(Gong et al., 2020; Veraart et al., 2018). Furthermore, the weak ΔRD contrast at an OGSE frequency of 54 Hz and the reduced ΔRD contrast at 28 Hz strongly suggested the lack of sensitivity to restricted neuronal structures at OGSE frequency lower than 50 Hz and b-value less than 1000 s/mm$^2$. Our study suggested that the minimal OGSE frequency to establish the association of ΔRD with axon diameter in the human brain *in vivo* should be larger than 54 Hz.

To achieve high OGSE frequencies at adequate b-values in human *in vivo*, simultaneously high gradient amplitude, slew rate, and PNS threshold are essential(Zhu et al., 2025). Human head-only high-performance gradients, such as the investigational MAGNUS systems used in this study, enable the potential translation of novel imaging contrast ΔRD from preclinical studies(Alderson et al., 2025) to human brain *in vivo*. In this study, a maximum OGSE frequency of 100 Hz was achieved at 280 mT/m gradient amplitude. MRI systems with maximum gradient amplitude higher than 300 mT/m may achieve higher OGSE frequency larger than 100 Hz(Zhu et al., 2025), which may further increase the sensitivity of ΔRD to brain microstructures at shorter length scale and/or axon diameters.

Evaluation of ΔRD in patients to correlate image contrast with specific tissue microstructural alterations in diseases is a promising new quantitative tool for assessment of neurologic diseases. In this study, ΔRD maps at OGSE 85 Hz showed distinct spatial patterns with reduced ΔRD in the corticospinal tract in a subject with a prior middle cerebral artery stroke. Although the image contrast may not be directly related to past medical history, the finding is intriguing and requires additional study for validation. For example, Zhou et al(Zhou et al., 2025) reported a higher ΔMD between OGSE $f$=40 Hz and PGSE Δ=40 ms in human acute ischemic lesions compared to contralateral white matter, with the largest effects in internal capsule and corona radiata. The altered ΔMD, ΔAD and ΔRD were indicated to be caused by neurite beading that is accompanied by both changed distribution of axonal diameter and increased volume fraction in Monte Carlo simulations(Zhou et al.). Future studies in dedicated patient populations will be performed to investigate potential tissue microstructural alterations.

Other time-dependent diffusion methods have been used to assess restricted diffusion and pore size, including the temporal diffusion ratio (TDR)(Dell'Acqua et al., 2019; Warner et al., 2023), a model-free method using PGSE diffusion encoding. Compared to the OGSE-based approach(Alderson et al., 2025; Harkins et al., 2021), TDR uses PGSE diffusion acquisitions at two relatively long diffusion times, ranging from 9 ms to 35 ms at a high b-value ≥6000 s/mm$^2$. A key similarity between TDR and the OGSE-based approach is the assumption that extra-axonal water diffusion is fast, Gaussian, and diffusion time-independent. This contribution can be suppressed either by using high b-values in TDR or, as in the present study, by subtracting RD measurements acquired at two diffusion times with a moderate b-value of 500 to 1000 s/mm$^2$. Therefore, both the temporal diffusion ratio from the two PGSE in TDR and ΔRD from OGSE-PGSE may reflect restricted diffusion within large axons. Direct comparison of the sensitivity of short-diffusion-time OGSE at moderate b-values (e.g., 1000-2000 s/mm$^2$) versus long-diffusion-time PGSE at high b-values for detecting intra-axonal restricted diffusion will be of considerable interest. Future scans of both PGSE-based and OGSE-based approaches will be performed.

Time-dependent diffusion MRI has also been shown to have high sensitivity to water exchange between intracellular space and extracellular space(Aggarwal et al., 2020; Chakwizira et al., 2023; Shi et al., 2024). Probing water exchange typically requires high-order diffusion measurements, such as diffusion kurtosis imaging at higher b-values ≥1500 s/mm$^2$. In this study, a maximum b-value of 1000 s/mm$^2$ was acquired to estimate diffusivity. Consequently, the sensitivity of ΔRD to water exchange in the white matter of healthy volunteers was minimized. Assessment of water exchange in white matter is clinically relevant, as it may provide insight into processes such as myelination and demyelination(Aggarwal et al., 2020). Future studies employing high b-value and optimized diffusion times will be performed to image water exchange in healthy volunteers and individuals with known demyelination lesions.

In this study, the maximum allowable mixing time in PGSE encoding was obtained based on TE, which is determined by the OGSE encoding in each time-dependent DTI acquisition. Therefore, the mixing time in PGSE encoding of each ΔRD measurement was different, ranging from 38 ms to 103 ms. Previous human brain *in vivo* studies(Fieremans et al., 2016; Johansson et al., 2024) have shown less than 0.05 μm$^2$/ms RD changes when the diffusion time ranged from 15 ms to 100 ms, at a b-value of 1000 s/mm$^2$. In future studies, protocol optimization, perhaps best pursued through simulation and signal modeling of tissue microstructures, will be considered not only for OGSE frequency, b-value, and TE, but also for PGSE mixing time.

Diffusivity in this study was measured from diffusion-weighted images at b-value of 500-1000 s/mm$^2$ relative to $T_2$-weighted images at b-value of 0, instead of a non-zero minimal b-value (e.g., 50 s/mm$^2$) by ignoring the effect of intravoxel incoherent motion (IVIM)(Le Bihan, 2019) on ΔRD. It was assumed that IVIM is diffusion time-independent in the range of OGSE 100 Hz to PGSE 103 ms used in this study, and subtracting RD at two diffusion times will eliminate the effect of IVIM on ΔRD. Future studies will be performed to evaluate the effect of IVIM on the image contrast of ΔRD.

This study has several limitations. First, the number of subjects was limited in this single-center study. Future studies in a large cohort of healthy subjects across multiple centers will be performed to better capture inter-subject variability and to investigate age- and sex-related effects on ∆RD(Tetreault et al., 2020), thereby establishing normative age- and sex-matched baseline values. Repeated scans will also be included to evaluate the repeatability of ∆RD measurements. Obviously, to be useful clinical studies of patients with known neurologic conditions such as stroke will need to be performed to validate the technique's ability to identify and monitor disease progression. Second, the association of ∆RD with neurite microstructure is not yet applicable to crossing fibers. For cross fibers, the measured RD could be close to the measured AD due to the contribution from diffusivity parallel to the axon. The expected diffusion time-dependent behavior of RD may not be applied to the crossing fibers. In future studies, the spherical mean techniques(Fan et al., 2020) will be applied to reduce sensitivity to fiber crossings and orientation dispersion. Furthermore, the current study did not include histological validation, and the relationship between ΔRD and axon diameter in human white matter remains inferred from simulation and preclinical data. Future work combining OGSE diffusion MRI with post-mortem tissue characterization will be needed to establish direct validation in humans.

In conclusion, this study demonstrates that time-dependent radial diffusivity measured with high OGSE frequencies on high-performance gradient MRI systems consistently revealed elevated ΔRD in the healthy corticospinal tract of the human brain *in vivo*. These observations suggest that ΔRD is sensitive to white matter microstructure associated with large axons in major white matter tracts, consistent with findings from preclinical studies. The baseline measurements established here provide a foundation for future evaluation of ΔRD as a non-invasive imaging biomarker of neuronal microstructural alterations in neurological disorders and diseases.

## Fundings

This work was supported by the National Institutes of Health (Grant No. U01 EB034313 and S10 OD030415).

## Declaration of Competing Interests

Dr. Zhu, Mr. McGovern, Dr. Yeo, and Dr. Li are currently employees in GE HealthCare.

## Acknowledgements

The authors thank Dr. Jia Guo from GE HealthCare for helping with the research pulse sequence modification.

The authors acknowledge the funding support from the National Institutes of Health (NIH) under award Number U01 EB034313, S10 OD030415. The content is solely the responsibility of the authors and does not necessarily represent the official views of the NIH.

The opinions and assertions expressed herein are those of the authors and do not necessarily reflect the official policy or position of the Uniformed Services University or the Department of War.

## References

Aggarwal, M., Jones, M. V., Calabresi, P. A., Mori, S., & Zhang, J. (2012). Probing mouse brain microstructure using oscillating gradient diffusion MRI. *Magn Reson Med*, *67*(1), 98-109. https://doi.org/10.1002/mrm.22981

Aggarwal, M., Smith, M. D., & Calabresi, P. A. (2020). Diffusion-time dependence of diffusional kurtosis in the mouse brain. *Magn Reson Med*, *84*(3), 1564-1578. https://doi.org/10.1002/mrm.28189

Alderson, H. E., Does, M. D., Hutchinson, E. B., & Harkins, K. D. (2025). Evaluation of diffusion time–dependent changes in radial diffusivity as a surrogate for axon diameter. *Magnetic resonance in medicine*, *94*(3), 1248-1256. https://doi.org/https://doi.org/10.1002/mrm.30538

Bhushan, C., Haldar, J. P., Choi, S., Joshi, A. A., Shattuck, D. W., & Leahy, R. M. (2015). Co-registration and distortion correction of diffusion and anatomical images based on inverse contrast normalization. *Neuroimage*, *115*, 269-280. https://doi.org/10.1016/j.neuroimage.2015.03.050

Bhushan, C., Haldar, J. P., Joshi, A. A., & Leahy, R. M. (2012). Correcting Susceptibility-Induced Distortion in Diffusion-Weighted MRI using Constrained Nonrigid Registration. *Signal and Information Processing Association Annual Summit and Conference (APSIPA), ... Asia-Pacific. Asia-Pacific Signal and Information Processing Association Annual Summit and Conference*, *2012*, http://ieeexplore.ieee.org/stamp/stamp.jsp?tp=&arnumber=6412009. https://www.ncbi.nlm.nih.gov/pubmed/26767197

Bin, J. M., Suminaite, D., Benito-Kwiecinski, S. K., Kegel, L., Rubio-Brotons, M., Early, J. J., Soong, D., Livesey, M. R., Poole, R. J., & Lyons, D. A. (2024). Importin 13-dependent axon diameter growth regulates conduction speeds along myelinated CNS axons. *Nature Communications*, *15*(1), 1790. https://doi.org/10.1038/s41467-024-45908-6

Chakwizira, A., Zhu, A., Foo, T., Westin, C. F., Szczepankiewicz, F., & Nilsson, M. (2023). Diffusion MRI with free gradient waveforms on a high-performance gradient system: Probing restriction and exchange in the human brain. *Neuroimage*, *283*, 120409. https://doi.org/10.1016/j.neuroimage.2023.120409

Chien, N., Cho, Y.-H., Wang, M.-Y., Tsai, L.-W., Yeh, C.-Y., Li, C.-W., Lan, P., Wang, X., Liu, K.-L., & Chang, Y.-C. (2025). Deep learning based multi-shot breast diffusion MRI: Improving imaging quality and reduced distortion. *European journal of radiology*, *193*. https://doi.org/10.1016/j.ejrad.2025.112419

Cochran, R. L., Bradley, W. R., Dhami, R. S., Milshteyn, E., Pohl, M., Ghosh, S., Nakrour, N., Lan, P., Wang, X., Guidon, A., & Harisinghani, M. G. (2026). Comparing prostate diffusion weighted images reconstructed with a commercial deep-learning product to a deep learning phase corrected model at 1.5 T. *Clinical Imaging*, *130*. https://doi.org/10.1016/j.clinimag.2025.110681

Dai, E., Zhu, A., Yang, G. K., Quah, K., Tan, E. T., Fiveland, E., Foo, T. K. F., & McNab, J. A. (2023). Frequency-dependent diffusion kurtosis imaging in the human brain using an oscillating gradient spin echo sequence and a high-performance head-only gradient. *Neuroimage*, *279*, 120328. https://doi.org/10.1016/j.neuroimage.2023.120328

Dell'Acqua, F., Dallyn, R., Chiappiniello, A., Beyh, A., Tax, C., Jones, D. K., & Catani, M. (2019). Temporal Diffusion Ratio (TDR): A Diffusion MRI technique to map the fraction and spatial distribution of large axons in the living human brain. 27th Annual Meeting of the ISMRM, Montreal, Canada,

Duval, T., Saliani, A., Nami, H., Nanci, A., Stikov, N., Leblond, H., & Cohen-Adad, J. (2018). Axons morphometry in the human spinal cord. *bioRxiv*, 282434. https://doi.org/10.1101/282434

Fan, Q., Nummenmaa, A., Witzel, T., Ohringer, N., Tian, Q., Setsompop, K., Klawiter, E. C., Rosen, B. R., Wald, L. L., & Huang, S. Y. (2020). Axon diameter index estimation independent of fiber orientation distribution using high-gradient diffusion MRI. *Neuroimage*, *222*, 117197. https://doi.org/10.1016/j.neuroimage.2020.117197

Feinberg, D. A., Beckett, A. J. S., Vu, A. T., Stockmann, J., Huber, L., Ma, S., Ahn, S., Setsompop, K., Cao, X., Park, S., Liu, C., Wald, L. L., Polimeni, J. R., Mareyam, A., Gruber, B., Stirnberg, R., Liao, C., Yacoub, E., Davids, M.,...Dietz, P. (2023). Next-generation MRI scanner designed for ultra-high-resolution human brain imaging at 7 Tesla. *Nat Methods*, *20*(12), 2048-2057. https://doi.org/10.1038/s41592-023-02068-7

Fieremans, E., Burcaw, L. M., Lee, H. H., Lemberskiy, G., Veraart, J., & Novikov, D. S. (2016). In vivo observation and biophysical interpretation of time-dependent diffusion in human white matter. *Neuroimage*, *129*, 414-427. https://doi.org/10.1016/j.neuroimage.2016.01.018

Foo, T. K., Tan, E. T., Vermilyea, M. E., Hua, Y., Fiveland, E. W., Piel, J. E., Park, K., Ricci, J., Thompson, P. S., & Graziani, D. (2020). Highly efficient head-only magnetic field insert gradient coil for achieving simultaneous high gradient amplitude and slew rate at 3.0 T (MAGNUS) for brain microstructure imaging. *Magnetic resonance in medicine*, *83*(6), 2356-2369.

Gong, T., Tong, Q., He, H., Sun, Y., Zhong, J., & Zhang, H. (2020). MTE-NODDI: Multi-TE NODDI for disentangling non-T2-weighted signal fractions from compartment-specific T2 relaxation times. *Neuroimage*, *217*, 116906. https://doi.org/10.1016/j.neuroimage.2020.116906

Harkins, K. D., Beaulieu, C., Xu, J., Gore, J. C., & Does, M. D. (2021). A simple estimate of axon size with diffusion MRI. *Neuroimage*, *227*, 117619. https://doi.org/10.1016/j.neuroimage.2020.117619

Holland, D., Kuperman, J. M., & Dale, A. M. (2010). Efficient correction of inhomogeneous static magnetic field-induced distortion in Echo Planar Imaging. *Neuroimage*, *50*(1), 175-183. https://doi.org/10.1016/j.neuroimage.2009.11.044

Huang, S. Y., Tian, Q., Fan, Q., Witzel, T., Wichtmann, B., McNab, J. A., Bireley, J. D., Machado, N., Klawiter, E. C., & Mekkaoui, C. (2020). High-gradient diffusion MRI reveals distinct estimates of axon diameter index within different white matter tracts in the in vivo human brain. *Brain Structure and Function*, *225*(4), 1277-1291.

Johansson, J., Lagerstrand, K., Björkman-Burtscher, I. M., Laesser, M., Hebelka, H., & Maier, S. E. (2024). Normal Brain and Brain Tumor ADC: Changes Resulting From Variation of Diffusion Time and/or Echo Time in Pulsed-Gradient Spin Echo Diffusion Imaging. *Investigative radiology*, *59*(10), 727-736. https://doi.org/10.1097/rli.0000000000001081

Lan, P., Huang, S. S., Bhushan, C., Wang, X., Lee, S. K., Yeo, D. T. B., Huang, R. Y., Maller, J. J., McNab, J. A., & Zhu, A. (2026). Human Brain High-Resolution Diffusion MRI With Optimized, Slice-By-Slice, Zeroth and First Order B0 Shimming in Head-Only High-Gradient MRI Systems. *Magn Reson Med*, *95*(6), 3322-3332. https://doi.org/10.1002/mrm.70310

Le Bihan, D. (2019). What can we see with IVIM MRI? *Neuroimage*, *187*, 56-67. https://doi.org/https://doi.org/10.1016/j.neuroimage.2017.12.062

Lebel, R. M. (2020). Performance characterization of a novel deep learning-based MR image reconstruction pipeline. *arXiv preprint arXiv:2008.06559*.

Li, C., Fieremans, E., Novikov, D. S., Ge, Y., & Zhang, J. (2023). Measuring water exchange on a preclinical MRI system using filter exchange and diffusion time dependent kurtosis imaging. *Magn Reson Med*, *89*(4), 1441-1455. https://doi.org/10.1002/mrm.29536

Ma, Y., Eskandarian, L., Chan, K.-S., Lee, H., Ramos-Llordén, G., Bhatt, A., Gerold, J., Mahmutovic, M., Keil, B., Lee, H.-H., & Huang, S. Y. (2026). Axon Diameter Mapping in the Living Human Brain with Ultra-High-Gradient Diffusion MRI at 500 mT/m Gradient Strength. *Human Brain Mapping*, *47*(8), e70553. https://doi.org/https://doi.org/10.1002/hbm.70553

Michael, E. S., Hennel, F., & Pruessmann, K. P. (2022). Evaluating diffusion dispersion across an extended range of b-values and frequencies: Exploiting gap-filled OGSE shapes, strong gradients, and spiral readouts. *Magn Reson Med*, *87*(6), 2710-2723. https://doi.org/10.1002/mrm.29161

Mori, S., Oishi, K., Jiang, H., Jiang, L., Li, X., Akhter, K., Hua, K., Faria, A. V., Mahmood, A., Woods, R., Toga, A. W., Pike, G. B., Neto, P. R., Evans, A., Zhang, J., Huang, H., Miller, M. I., van Zijl, P., & Mazziotta, J. (2008). Stereotaxic white matter atlas based on diffusion tensor imaging in an ICBM template. *Neuroimage*, *40*(2), 570-582. https://doi.org/https://doi.org/10.1016/j.neuroimage.2007.12.035

Olesen, J. L., Ostergaard, L., Shemesh, N., & Jespersen, S. N. (2022). Diffusion time dependence, power-law scaling, and exchange in gray matter. *Neuroimage*, *251*, 118976. https://doi.org/10.1016/j.neuroimage.2022.118976

Ramos-Llordén, G., Lee, H.-H., Davids, M., Dietz, P., Krug, A., Kirsch, J. E., Mahmutovic, M., Müller, A., Ma, Y., Lee, H., Maffei, C., Yendiki, A., Bilgic, B., Park, D. J., Tian, Q., Clifford, B., Lo, W.-C., Stocker, S., Fischer, J.,...Huang, S. Y. (2025). Ultra-high gradient connectomics and microstructure MRI scanner for imaging of human brain circuits across scales. *Nature Biomedical Engineering*. https://doi.org/10.1038/s41551-025-01457-x

Rosler, M. B., Leussler, C., Brunner, D. O., Schmid, T., Hennel, F., Luechinger, R., Weiger, M., & Pruessmann, K. P. (2020). A transmit-receive array for brain imaging with a high-performance gradient insert. *Magn Reson Med*, *84*(4), 2278-2289. https://doi.org/10.1002/mrm.28261

Seidl, A. H. (2014). Regulation of conduction time along axons. *Neuroscience*, *276*, 126-134. https://doi.org/10.1016/j.neuroscience.2013.06.047

Shi, D., Liu, F., Li, S., Chen, L., Jiang, X., Gore, J. C., Zheng, Q., Guo, H., & Xu, J. (2024). Restriction-induced time-dependent transcytolemmal water exchange: Revisiting the Kärger exchange model. *Journal of magnetic resonance*, *367*, 107760. https://doi.org/https://doi.org/10.1016/j.jmr.2024.107760

Tan, E. T., Marinelli, L., Slavens, Z. W., King, K. F., & Hardy, C. J. (2013). Improved correction for gradient nonlinearity effects in diffusion-weighted imaging. *J Magn Reson Imaging*, *38*(2), 448-453. https://doi.org/10.1002/jmri.23942

Tetreault, P., Harkins, K. D., Baron, C. A., Stobbe, R., Does, M. D., & Beaulieu, C. (2020). Diffusion time dependency along the human corpus callosum and exploration of age and sex differences as assessed by oscillating gradient spin-echo diffusion tensor imaging. *Neuroimage*, *210*, 116533. https://doi.org/10.1016/j.neuroimage.2020.116533

Van, A. T., Holdsworth, S. J., & Bammer, R. (2014). In vivo investigation of restricted diffusion in the human brain with optimized oscillating diffusion gradient encoding. *Magn Reson Med*, *71*(1), 83-94. https://doi.org/10.1002/mrm.24632

Veraart, J., Novikov, D. S., & Fieremans, E. (2018). TE dependent Diffusion Imaging (TEdDI) distinguishes between compartmental T(2) relaxation times. *Neuroimage*, *182*, 360-369. https://doi.org/10.1016/j.neuroimage.2017.09.030

Warner, W., Palombo, M., Cruz, R., Callaghan, R., Shemesh, N., Jones, D. K., Dell'Acqua, F., Ianus, A., & Drobnjak, I. (2023). Temporal Diffusion Ratio (TDR) for imaging restricted diffusion: Optimisation and pre-clinical demonstration. *Neuroimage*, *269*, 119930. https://doi.org/https://doi.org/10.1016/j.neuroimage.2023.119930

Wu, D., Martin, L. J., Northington, F. J., & Zhang, J. (2014). Oscillating gradient diffusion MRI reveals unique microstructural information in normal and hypoxia-ischemia injured mouse brains. *Magn Reson Med*, *72*(5), 1366-1374. https://doi.org/10.1002/mrm.25441

Wu, D., Zhang, Y., Cheng, B., Mori, S., Reeves, R. H., & Gao, F. J. (2021). Time-dependent diffusion MRI probes cerebellar microstructural alterations in a mouse model of Down syndrome. *Brain Commun*, *3*(2), fcab062. https://doi.org/10.1093/braincomms/fcab062

Xu, J., Jiang, X., Devan, S. P., Arlinghaus, L. R., McKinley, E. T., Xie, J., Zu, Z., Wang, Q., Chakravarthy, A. B., Wang, Y., & Gore, J. C. (2021). MRI-cytometry: Mapping nonparametric cell size distributions using diffusion MRI. *Magn Reson Med*, *85*(2), 748-761. https://doi.org/10.1002/mrm.28454

Xu, J., Xie, J., Semmineh, N. B., Devan, S. P., Jiang, X., & Gore, J. C. (2023). Diffusion time dependency of extracellular diffusion. *Magn Reson Med*, *89*(6), 2432-2440. https://doi.org/10.1002/mrm.29594

Zhou, M., Stobbe, R., Budde, M., Buck, B., Kate, M., Lloret, M., Fairall, P., Butcher, K., Shuaib, A., Emery, D., Feiweier, T., & Beaulieu, C. Oscillating gradient spin echo diffusion time effects implicate variations in neurite beading for the heterogeneous reduced diffusion in human acute ischemic stroke lesions. *Magnetic resonance in medicine*, *n/a*(n/a). https://doi.org/https://doi.org/10.1002/mrm.30618

Zhu, A., Michael, E. S., Li, H., Sprenger, T., Hua, Y., Lee, S. K., Yeo, D. T. B., McNab, J. A., Hennel, F., Fieremans, E., Wu, D., Foo, T. K. F., & Novikov, D. S. (2025). Engineering clinical translation of OGSE diffusion MRI. *Magn Reson Med*, *94*(3), 913-936. https://doi.org/10.1002/mrm.30510

Zhu, A., Shih, R., Huang, R. Y., DeMarco, J. K., Bhushan, C., Morris, H. D., Kohls, G., Yeo, D. T. B., Marinelli, L., Mitra, J., Hood, M., Ho, V. B., & Foo, T. K. F. (2023). Revealing tumor microstructure with oscillating diffusion encoding MRI in pre-surgical and post-treatment glioma patients. *Magn Reson Med*, *90*(5), 1789-1801. https://doi.org/10.1002/mrm.29758

## Figures and Tables

***Table 1.*** *Imaging parameters of time-dependent DTI scans.*

| | **Td-DTI High-frequency Medium-b Long-TE**ˣ | **Td-DTI Medium-frequency High-b Long-TE*** | **Td-DTI Medium-frequency Low-b Short-TE**^ | **Td-DTI Low-frequency High-b Short-TE‡** | **Td-DTI Low-frequency Low-b Long-TE†** |
|---|---|---|---|---|---|
| Imaging plane | Axial | | | | |
| Field of view | 22 cm | | | | |
| Spatial resolution | 2 x 2 x 2 $mm^3$ | | | | |
| # of slices | 64 | | | | |
| Acceleration | No acceleration | | | | |
| Echo spacing | 364 $\mu$s | | | | |
| Partial ky sampling | 65% with 16 oversampled ky lines | | | | |
| TE | 121 ms | 115 ms | 68 ms | 58 ms | 92 ms |
| b-value | 800 s/$mm^2$ | 1000 s/$mm^2$ | 500 s/$mm^2$ | 1000 s/$mm^2$ | 500 s/$mm^2$ |
| Diffusion time | OG: 100 Hz<br>PG: $\Delta$=103 ms, $\delta$=1.7 ms | OG: 85 Hz<br>PG: $\Delta$=96 ms, $\delta$=2.2 ms | OG: 85 Hz<br>PG: $\Delta$=49 ms, $\delta$=2.1 ms | OG: 54 Hz<br>PG: $\Delta$=38 ms, $\delta$=4.0 ms | OG: 28 Hz<br>PG: $\Delta$=68 ms, $\delta$=8.0 ms |
| Gradient amplitude of diffusion encoding | OG: 280 mT/m<br>PG: 191 mT/m | OG: 264 mT/m<br>PG: 177 mT/m | OG: 279 mT/m<br>PG: 181 mT/m | OG: 272 mT/m<br>PG: 171 mT/m | OG: 65 mT/m<br>PG: 41 mT/m |
| TR | 15000 ms | 15000 ms | 8200 ms | 7800 ms | 7800 ms |
| # of diffusion directions | 20 | 20 | 20 | 20 | 20 |
| # of b=0 images | 2 | 2 | 2 | 2 | 2 |
| Scan time | 11 mins | 11 mins | 6 mins | 5 mins 43 secs | 5 mins 43 secs |

OG: oscillating gradient; PG: pulsed gradient.

ˣ Highest OGSE frequency and adequate b-value at TE≤125 ms

* Highest OGSE frequency at b≥1000 s/$mm^2$ and TE≤120 ms

^ Lower b-value of 500 s/$mm^2$ at the same OGSE frequency as in *

‡ Shortest TE at an OGSE frequency at b≥1000 s/$mm^2$

† Highest OGSE frequency at b-value ≥500 s/$mm^2$ and TE ≤100 ms in whole-body MRI system at 80 mT/m and 200 T/m/s

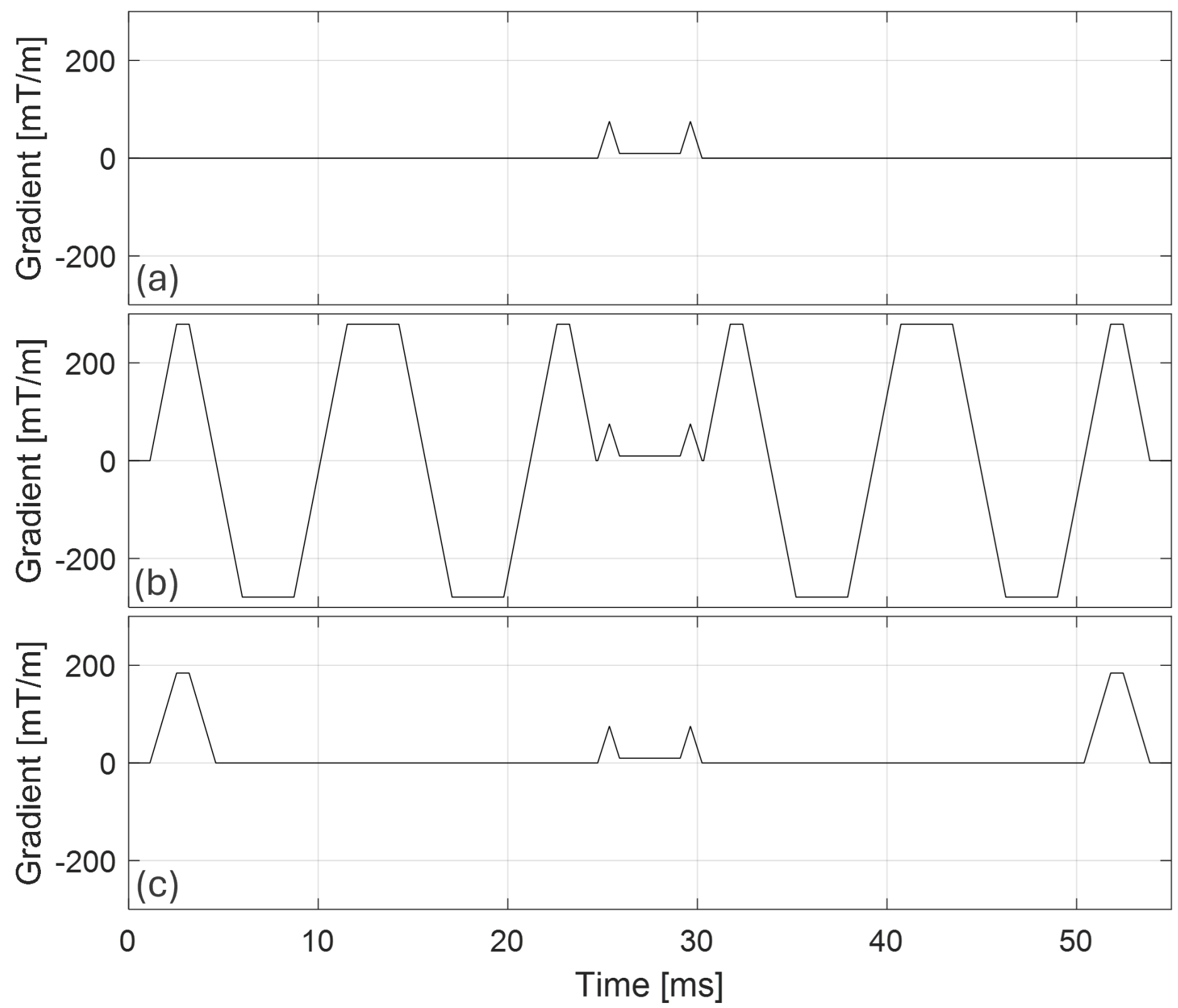

***Figure 1**. The gradient waveforms of $T_2$-weighted EPI (i.e., b=0 s/mm$^2$) (a), OGSE encoding (b), and PGSE encoding (c) used in this study. The crusher gradients and slice selection gradients of 180° RF pulse are also shown. The first and last gradient lobes of the OGSE encoding gradients are used as PGSE encoding gradients. Therefore, the maximum allowable mixing time in PGSE encoding was obtained.*

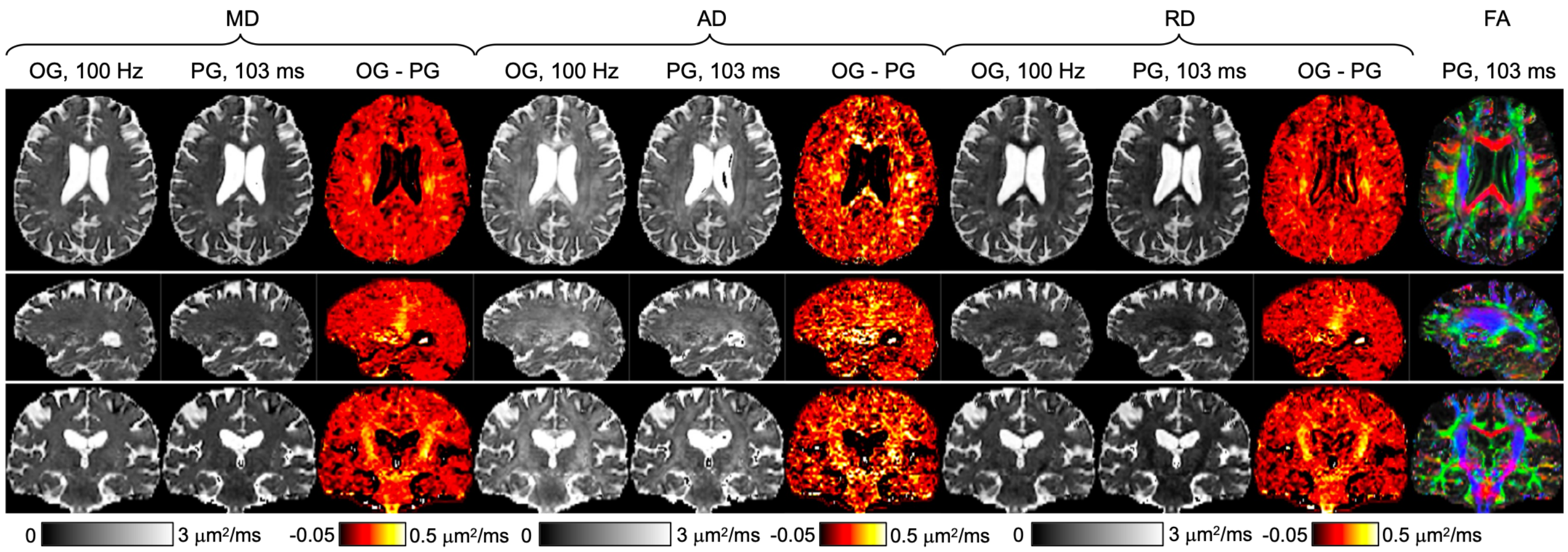

***Figure 2**. Axial (top row), sagittal (middle row), and coronal (bottom row) views of representative MD/AD/RD maps measured from OGSE 100 Hz encoding and PGSE encoding, as well as ΔMD/ΔAD/ΔRD, from the "High-frequency Medium-b Long-TE" protocol in a single subject ("Subject #1"). FA measured from the PGSE encoding is also shown.*

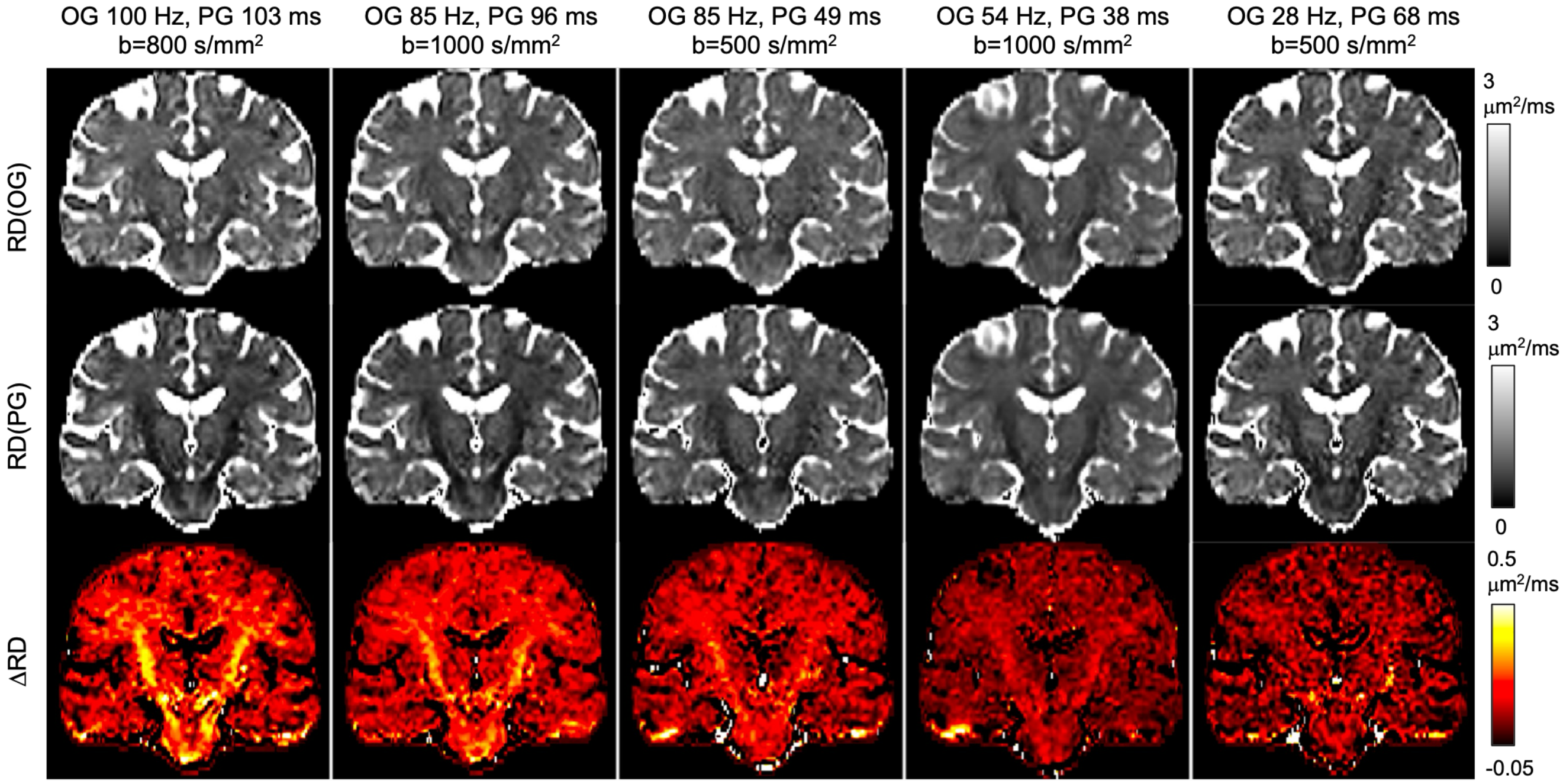

***Figure 3**. Coronal view of representative RD and ΔRD maps measured from OGSE encoding and PGSE encoding from all five time-dependent DTI protocols in a single subject ("Subject #2").*

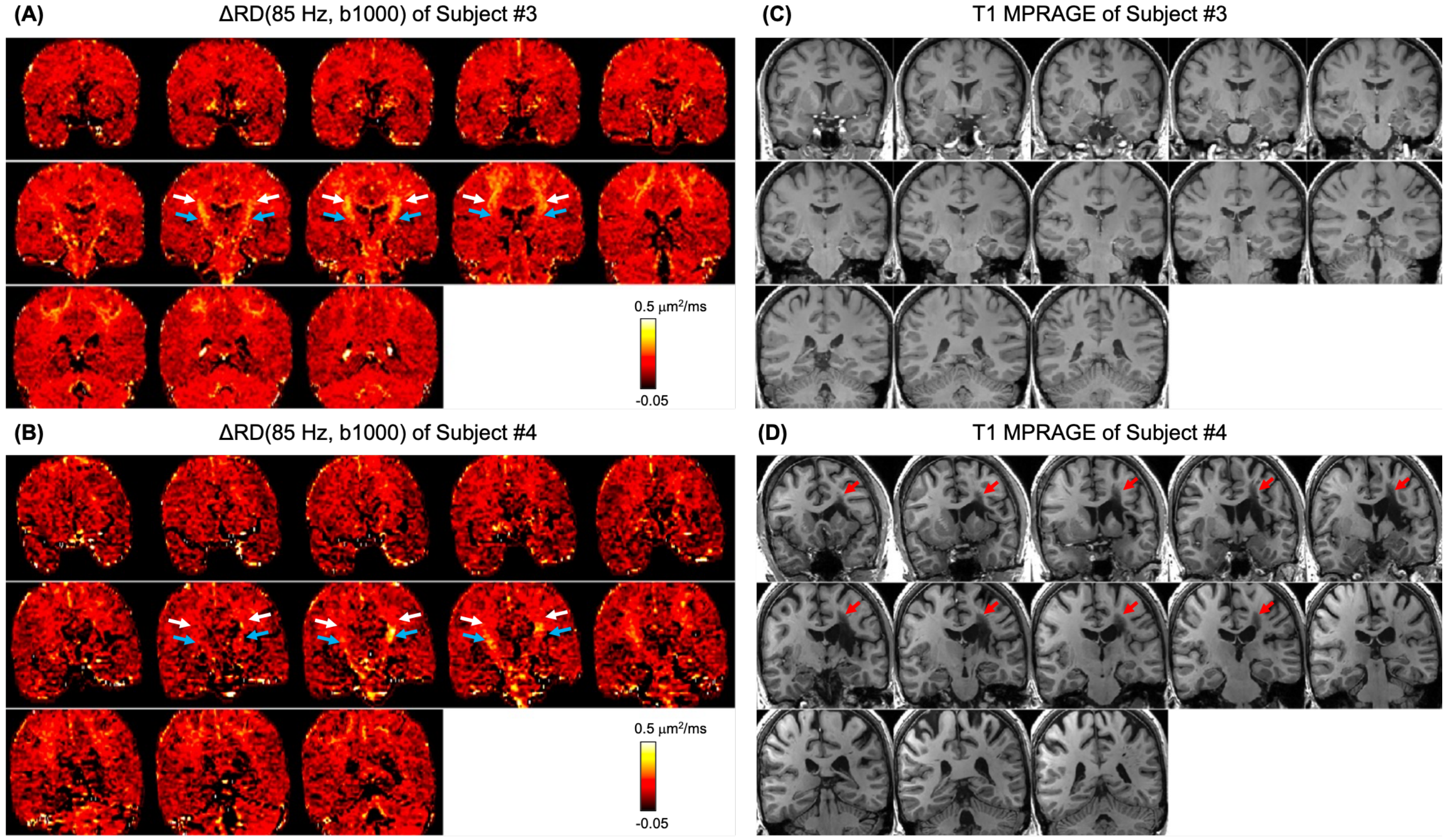

***Figure 4**. ΔRD(85 Hz, b1000) and $T_1$-weighted MPRAGE of multiple slices in two female subjects at similar ages. Hyperintense ΔRD(85 Hz, b1000) was observed in the white matter fibers of the major motor pathways, including corona radiata (white arrows in A) and internal capsule (blue arrows in A) of Subject #3 (A and C). Hypo-intensities in $T_1$-weighted MPRAGE were present in the left hemisphere of Subject #4 (red arrows in D). Subject #4 previously had a middle cerebral artery stroke. The corona radiata (white arrows in B) and internal capsule (blue arrows in B) exhibited smaller regions with hyperintense ΔRD(85 Hz, b1000) in Subject #4.*

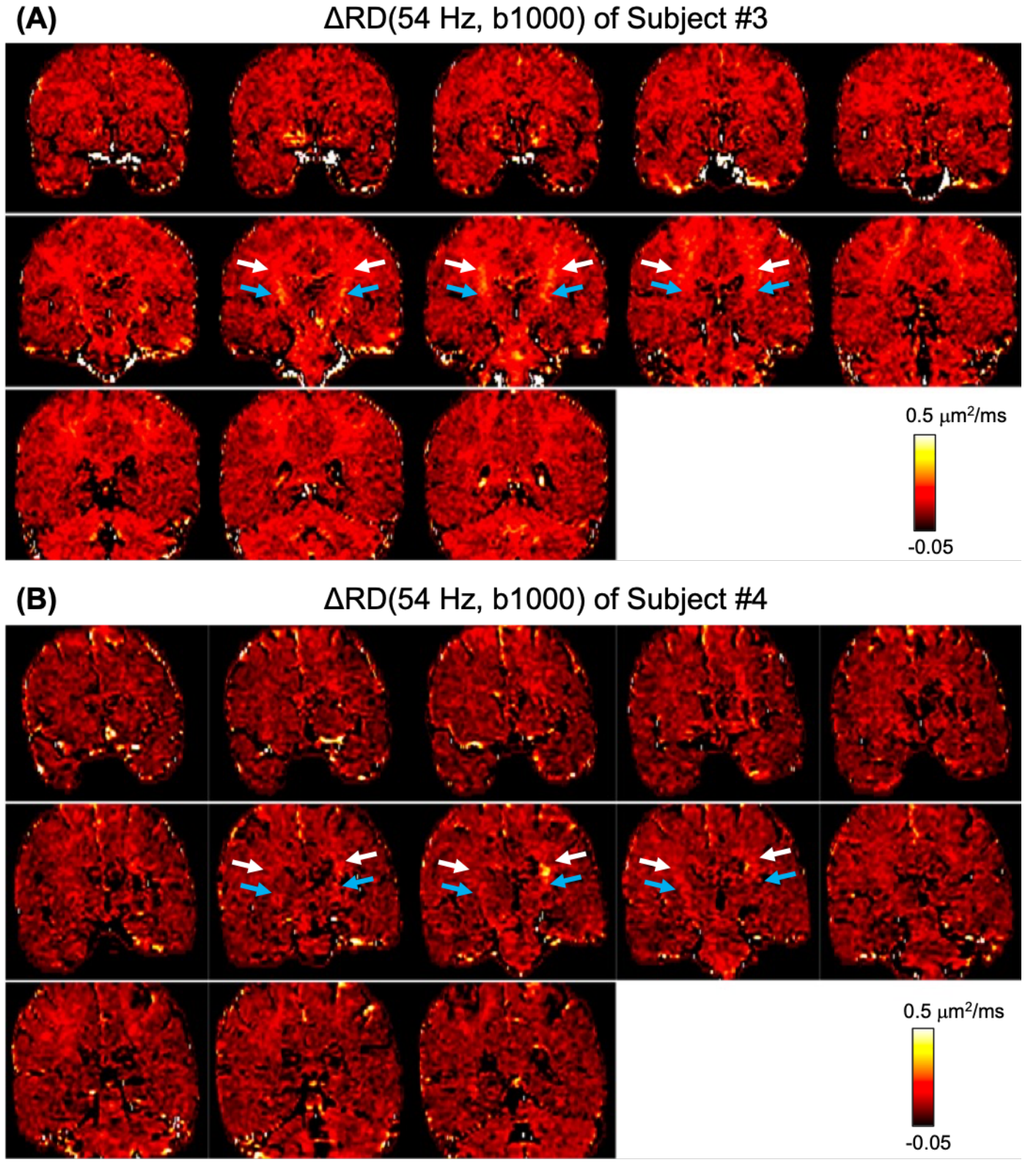

***Figure 5**. ΔRD(54 Hz, b1000) of multiple slices in the two subjects shown in Figure 4, i.e., Subject #3 (A) and Subject #4 (B). ΔRD(54 Hz, b1000) of the corona radiata (white arrows in A) and internal capsule (blue arrows in A) exhibited subtle hyper-intensities in Subject #3. ΔRD(54 Hz, b1000) of the internal capsule (blue arrows in B) exhibited reduced regions with hyperintense ΔRD in Subject #4.*

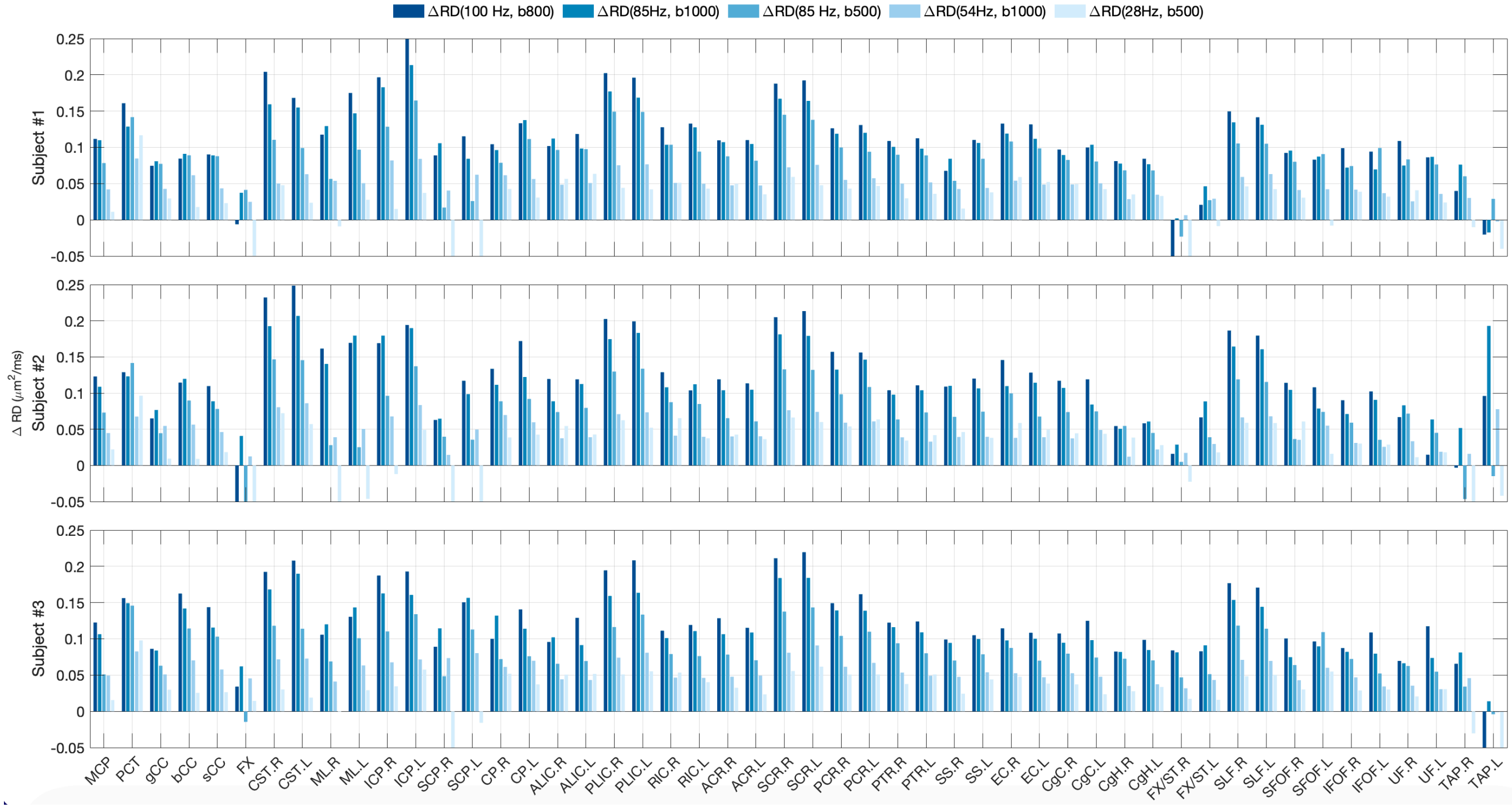

***Figure 6**. Mean ΔRD of each white matter parcel in Subject #1, Subject #2, and Subject #3. ΔRD measured from the five time-dependent DTI protocols is shown for each white matter parcel.*

***Table 2**. ΔRD measurements in the white matter parcels associated with the corticospinal tract, forceps minors and forceps majors. Parcels from the left (L) and right (R) hemispheres are shown separately. Results are presented for each healthy subject (Subjects #1, #2, #3), along with the mean ΔRD across all subjects. The unit of ΔRD is μm²/ms.*

| | | **ΔRD(100 Hz, b800)** | **ΔRD(85Hz, b1000)** | **ΔRD(85Hz, b500)** | **ΔRD(54Hz, b1000)** | **ΔRD(28 Hz, b500)** |
|---|---|---|---|---|---|---|
| Corticospinal tract | Atlas-defined Corticospinal tract ROI (CST.L / CST.R) | L: 0.168, 0.248, 0.208<br>R: 0.204, 0.232, 0.192 | L: 0.155, 0.207, 0.190<br>R: 0.160, 0.193, 0.168 | L: 0.099, 0.146, 0.114<br>R: 0.110, 0.147, 0.118 | L: 0.063, 0.086, 0.073<br>R: 0.050, 0.080, 0.072 | L: 0.024, 0.057, 0.019<br>R: 0.048, 0.072, 0.030 |
| | | L: 0.204 / R: 0.213 | L: 0.183 / R: 0.174 | L: 0.128 / R: 0.129 | L: 0.079 / R: 0.072 | L: 0.039 / R: 0.051 |
| | Superior corona radiata (SCR.L / SCR.R) | L: 0.192, 0.213, 0.220<br>R: 0.188, 0.205, 0.211 | L: 0.164, 0.179, 0.184<br>R: 0.167, 0.181, 0.184 | L: 0.138, 0.132, 0.143<br>R: 0.145, 0.133, 0.138 | L: 0.076, 0.074, 0.091<br>R: 0.072, 0.076, 0.081 | L: 0.048, 0.060, 0.062<br>R: 0.059, 0.066, 0.056 |
| | | L: 0.203 / R: 0.197 | L: 0.174 / R: 0.173 | L: 0.137 / R: 0.136 | L: 0.079 / R: 0.075 | L: 0.058 / R: 0.062 |
| | Posterior corona radiata (PCR.L / PCR.R) | L: 0.131, 0.156, 0.162<br>R: 0.126, 0.157, 0.149 | L: 0.120, 0.146, 0.139<br>R: 0.119, 0.133, 0.139 | L: 0.094, 0.108, 0.110<br>R: 0.100, 0.098, 0.104 | L: 0.057, 0.061, 0.067<br>R: 0.055, 0.059, 0.061 | L: 0.047, 0.064, 0.051<br>R: 0.043, 0.054, 0.051 |
| | | L: 0.150 / R: 0.143 | L: 0.134 / R: 0.131 | L: 0.102 / R: 0.102 | L: 0.062 / R: 0.058 | L: 0.054 / R: 0.050 |
| | Posterior limb of the internal capsule (PLIC.L / PLIC.R) | L: 0.196, 0.199, 0.208<br>R: 0.202, 0.203, 0.195 | L: 0.169, 0.183, 0.164<br>R: 0.177, 0.175, 0.159 | L: 0.149, 0.134, 0.134<br>R: 0.149, 0.130, 0.116 | L: 0.077, 0.073, 0.081<br>R: 0.075, 0.071, 0.074 | L: 0.042, 0.052, 0.055<br>R: 0.044, 0.062, 0.051 |
| | | L: 0.207 / R: 0.196 | L: 0.174 / R: 0.169 | L: 0.143 / R: 0.131 | L: 0.077 / R: 0.073 | L: 0.051 / R: 0.054 |
| | Cerebral Peduncle (CP.L / CP.R) | L: 0.133, 0.172, 0.141<br>R: 0.104, 0.134, 0.100 | L: 0.138, 0.122, 0.114<br>R: 0.096, 0.112, 0.132 | L: 0.112, 0.092, 0.076<br>R: 0.079, 0.089, 0.072 | L: 0.056, 0.060, 0.070<br>R: 0.061, 0.070, 0.061 | L: 0.031, 0.043, 0.037<br>R: 0.043, 0.039, 0.052 |
| | | L: 0.170 / R: 0.142 | L: 0.134 / R: 0.136 | L: 0.105 / R: 0.106 | L: 0.064 / R: 0.069 | L: 0.041 / R: 0.061 |
| | Pontine crossing tract (PCT) | 0.161, 0.129, 0.156 | 0.129, 0.123, 0.149 | 0.142, 0.142, 0.146 | 0.085, 0.068, 0.083 | 0.117, 0.097, 0.098 |
| | | 0.148 | 0.135 | 0.142 | 0.078 | 0.105 |
| Forceps minors | Genu of corpus callosum (gCC) | 0.074, 0.065, 0.086 | 0.081, 0.077, 0.084 | 0.077, 0.045, 0.063 | 0.043, 0.055, 0.051 | 0.029, 0.009, 0.030 |
| | | 0.084 | 0.086 | 0.070 | 0.048 | 0.029 |
| | Anterior corona radiata (ACR.L / ACR.R) | L: 0.110, 0.114, 0.115<br>R: 0.110, 0.119, 0.129 | L: 0.105, 0.105, 0.109<br>R: 0.107, 0.104, 0.106 | L: 0.082, 0.061, 0.070<br>R: 0.088, 0.065, 0.078 | L: 0.047, 0.040, 0.049<br>R: 0.048, 0.040, 0.048 | L: 0.035, 0.036, 0.023<br>R: 0.050, 0.043, 0.033 |
| | | L: 0.115 / R: 0.121 | L: 0.109 / R: 0.107 | L: 0.073 / R: 0.079 | L: 0.048 / R: 0.047 | L: 0.031 / R: 0.043 |
| Forceps majors | Splenium of corpus callosum (sCC) | 0.090, 0.110, 0.144 | 0.089, 0.089, 0.116 | 0.088, 0.078, 0.103 | 0.043, 0.046, 0.058 | 0.023, 0.018, 0.026 |
| | | 0.122 | 0.103 | 0.094 | 0.050 | 0.029 |
| | Tapetum (TAP.L / TAP.R) | L: -0.020, 0.096, -0.062<br>R: 0.040, -0.003, 0.066 | L: -0.017, 0.193, 0.014<br>R: 0.076, 0.052, 0.081 | L: 0.029, -0.015, -0.004<br>R: 0.060, -0.047, 0.034 | L: -0.002, 0.078, -0.001<br>R: 0.030, 0.016, 0.046 | L: -0.040, -0.042, -0.078<br>R: -0.010, -0.158, -0.030 |
| | | L: 0.047 / R: 0.082 | L: 0.077 / R: 0.093 | L: 0.048 / R: 0.068 | L: 0.039 / R: 0.045 | L: -0.009 / R: 0.003 |
| | Sagittal stratum (SS.L / SS.R) | L: 0.110, 0.120, 0.105<br>R: 0.068, 0.109, 0.099 | L: 0.106, 0.106, 0.100<br>R: 0.084, 0.110, 0.094 | L: 0.084, 0.074, 0.079<br>R: 0.054, 0.067, 0.070 | L: 0.044, 0.040, 0.054<br>R: 0.043, 0.039, 0.048 | L: 0.038, 0.038, 0.044<br>R: 0.016, 0.046, 0.024 |
| | | L: 0.115 / R: 0.099 | L: 0.108 / R: 0.095 | L: 0.079 / R: 0.072 | L: 0.046 / R: 0.044 | L: 0.039 / R: 0.036 |
| | Posterior thalamic radiation (PTR.L / PTR.R) | L: 0.113, 0.111, 0.124<br>R: 0.109, 0.104, 0.122 | L: 0.098, 0.104, 0.109<br>R: 0.101, 0.098, 0.116 | L: 0.089, 0.073, 0.080<br>R: 0.090, 0.064, 0.094 | L: 0.052, 0.033, 0.049<br>R: 0.050, 0.039, 0.053 | L: 0.036, 0.042, 0.051<br>R: 0.030, 0.034, 0.038 |
| | | L: 0.121 / R: 0.119 | L: 0.107 / R: 0.108 | L: 0.081 / R: 0.086 | L: 0.047 / R: 0.050 | L: 0.045 / R: 0.037 |

All-subject mean calculated by averaging ΔRD of all voxels in the white matter parcel from all three healthy subjects. L: left; R: right.

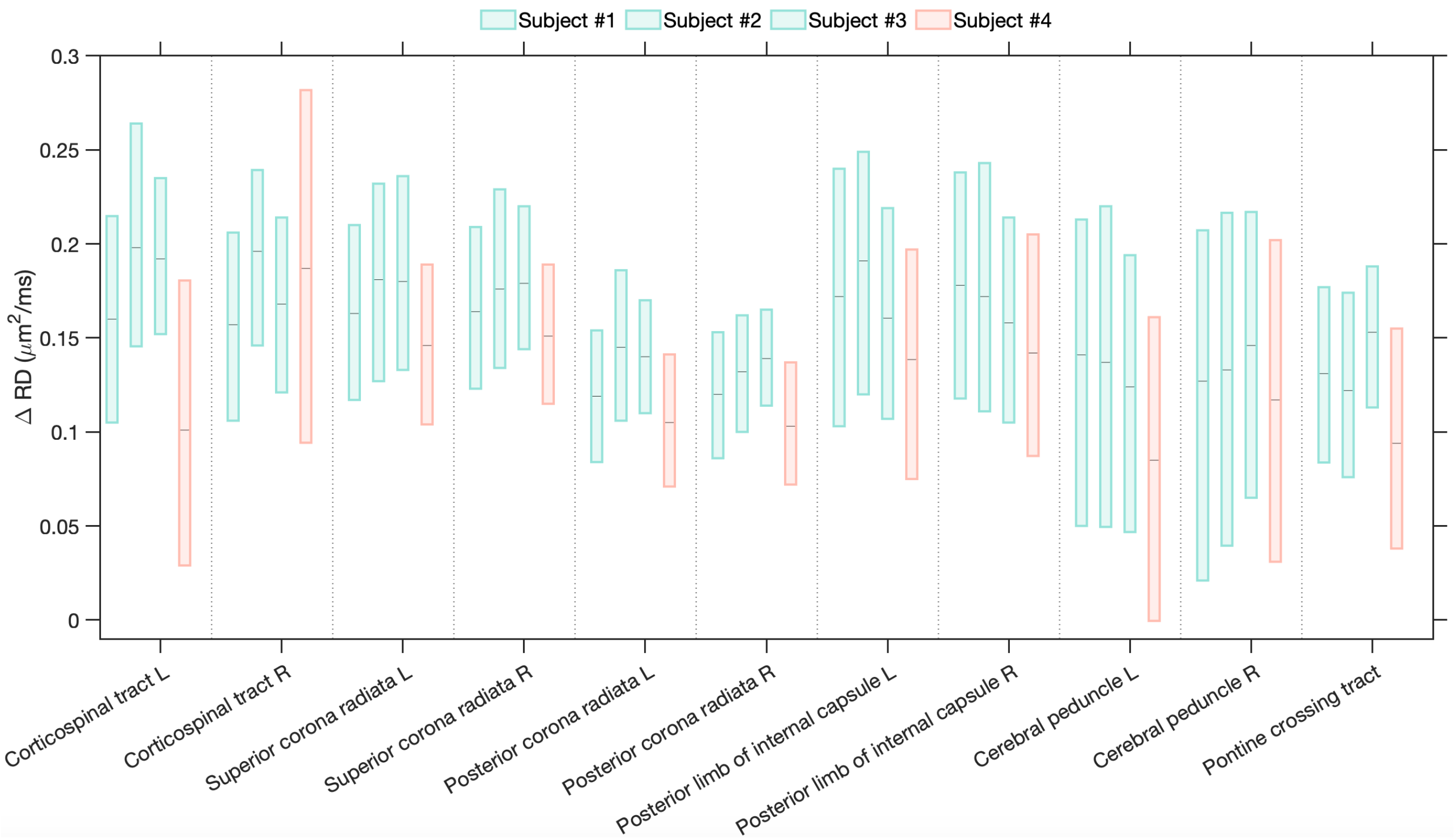

***Figure 7**. Box plots of ΔRD(85 Hz, b1000) in each white matter parcel of the corticospinal tract in Subjects #1, #2, and #3 (colored in light teal) versus Subject #4 (colored in light coral). The central mark of each box is the median, and the edges are the 25$^{th}$ and 75$^{th}$ percentiles.*